\documentclass[aps,epsf,amsfonts,floats,twocolumn,amssymb,amsmath,groupedaddress,showpacs,floatfix,nofootinbib]{revtex4}
\usepackage{graphicx}
\usepackage{epsf}

\usepackage[]{latexsym}
\usepackage{bm}
\usepackage{amsmath}

\newcommand{\be}{\begin{equation}}\newcommand{\ee}{\end{equation}}
\newcommand{\bea}{\begin{eqnarray}}\newcommand{\eea}{\end{eqnarray}}
\newcommand{\brr}{\begin{array}}\newcommand{\err}{\end{array}}
\newcommand{\bit}{\begin{itemize}}\newcommand{\eit}{\end{itemize}}
\newcommand{\ben}{\begin{enumerate}}\newcommand{\een}{\end{enumerate}}

\newcommand{\ba}{\begin{array}}
\newcommand{\ea}{\end{array}}

\def\1{{_{1}}}\def\2{{_{2}}}

\def\noHe0{:\;\!\!\;\!\!:H_e(0):\;\!\!\;\!\!:}
\def\noHm0{:\;\!\!\;\!\!:H_\mu(0):\;\!\!\;\!\!:}

\def\1{{_{1}}}\def\2{{_{2}}}

\begin{document}

\title{A hydrodynamic model for cooperating solidary countries}

\author{Roberto De Luca$^1$}

\author{Marco Di Mauro$^1$}

\author{Angelo Falzarano$^2$}

\author{Adele Naddeo$^3$}

\affiliation{ $^1$ Dipartimento di Fisica E.R.Caianiello, Universit\'a di Salerno, Fisciano (SA) - 84084, Italy}

\affiliation{ $^2$ Dipartimento di Scienze Economiche e Statistiche, Universit\'a di Napoli Federico II, Napoli - 80126, Italy}

\affiliation{ $^3$ INFN Sezione di Napoli, Napoli - 80126, Italy}


\begin{abstract}
The goal of international trade theories is to explain the exchange of goods and services between different countries, aiming to benefit from it. Albeit the idea is very simple and known since ancient history, smart policy and business strategies need to be implemented by each subject, resulting in a complex as well as not obvious interplay. In order to understand such a complexity, different theories have been developed since the sixteenth century and today new ideas still continue to enter the game. Among them, the so called classical theories are country-based and range from Absolute and Comparative Advantage theories by A. Smith and D. Ricardo to Factor Proportions theory by E. Heckscher and B. Ohlin. In this work we build a simple hydrodynamic model, able to reproduce the main conclusions of Comparative Advantage theory in its simplest setup, i.e. a two-country world with country $A$ and country $B$ exchanging two goods within a genuine exchange-based economy and a trade flow ruled only by market forces. The model is further generalized by introducing money in order to discuss its role in shaping trade patterns. Advantages and drawbacks of the model are also discussed together with perspectives for its improvement.
\end{abstract}

\maketitle

\section{Introduction}

International trade theories help to elucidate the basis for trade and the gains from trade, but also, how the gains from trade are generated, and  how are they large and how are they divided among the trading nations. Theories also explain the pattern of trade. That is, which commodities are traded and which are exported and imported by each country. Presumably a nation will voluntarily engage in trade only if it gains from import-export. Although the idea is very simple and known since ancient times, the policies and intelligent strategies adopted by different countries move on dynamics not always interpretable preventively. Even more so today, indeed, where the global trade is governed by multiple and changing factors, often combined in a subtle way and often demanding an explanation.

To understand this complex scenario, the approaches are different with different results since the sixteenth century.  The Leitmotiv of these contributions is the law of comparative advantage \cite{chipman}, which is still one of the most important and unchallenged laws of economics, with many practical applications to nations, as well as to individuals. The Comparative Advantage (CA) is also studied in terms of the opportunity cost theory, as reflected in production possibility frontiers or transformation curves.

Smith's theory \cite{smith1,smith2} for the first time considered the ability of a country to a better efficiency in the production of a good with respect to a second country. In a simple but ideal two-country world, that would result in an advantage for the first country, which could specialize in producing that good. At the same time, if the second country would be more efficient in producing another good, it could specialize in this effort. In this way a better efficiency makes production to increase and trade to flow according only to market forces with a clear benefit for people in both countries. In conclusion, according to Smith's theory both countries gain from exchange and their wealth can be measured in terms of the living standards of the local population. The main drawback in the above reasoning was essentially due to the possible occurrence of one country in the network without any absolute advantage in producing a good and the other efficient in producing both goods. This issue found a satisfying solution in the CA theory by Ricardo \cite{ricardo1}, whose main feature was the international immobility of production's factors. Indeed factors were considered as completely immobile among countries, while goods showed a complete mobility and a constant unit cost. In this context a comparative advantage takes place if a country can be more efficient in producing a particular good with respect to other goods. As a consequence, it will specialize in the production of that good and this makes international trade to depend on a difference in the comparative cost of producing goods. The net result is that each country should produce and export the goods in which it has a comparative advantage while importing those goods in which it has a comparative disadvantage within an environment characterized by open and free markets. This is clearly a drawback, which was addressed by Heckscher and Ohlin \cite{ohlin} theory. According to such a theory, countries differ with respect to production factors, such as land, labor and capital, whose cost depends on supply and demand. In this framework, a country specializes in the production and export of goods characterized by a great supply and cheaper production factors, while establishes to import goods being in short supply. Another drawback of CA theory is that it provides a static framework; in this respect one may wonder how trade patters change as a function of time. To this and other issues, such as the possibility to introduce further factors in the understanding of international trade flows, are devoted modern trade theories \cite{modern1}. As pointed out by Helpman \cite{helpman1}\cite{modern2} and Krugman \cite{krugman1}\cite{krugman2}\cite{krugman3}, modern theories have been developed in order to take into account mainly the increasing of the trade to income ratio, the concentration of trade flows as well as the expansion of intra-industry trade among industrialized countries. According to Markusen \cite{markusen1}\cite{markusen2} the expansion of intra-industry trade could be ascribed to the increasing of the demand for differentiated products with respect to that for homogeneous ones, while a model developed by Spence \cite{spence} and Dixit and Stiglitz \cite{dixit} focused on the link between trade and consumption. In this way policy changes may be recognized as the driving force behind the increase in trade volume. Finally the meaning and the role of CA idea in monetary economies has been discussed as well, together with its effect on the long-run equilibrium pattern of trade \cite{modern1}\cite{rmoney}\cite{krugman3bis}. In this context a direct impact of exchange rates changes on trade balance has been found and its effect on exchange rate policy widely investigated \cite{btrade1}\cite{btrade2}\cite{btrade3}\cite{btrade4}.

In summary, Ricardo's ideas represented a fundamental starting point for the development of modern theories of international trade, even if it is clear that the matter is really complex and characterized by networks of more than two countries and two goods to exchange and by the interplay of a lot of different variables. Indeed Comparative Advantage may still provide the underlying idea for the optimal allocation of any country's resources and the maximization of world welfare, and the consequence would be that the benefits of free trade outweigh the costs \cite{krugman4}, even if today it is believed that international trade couldn't find a complete explanation within a single theory but it may require resorting to different ones at the same time.

Among the physical models that can be employed to describe this kind of situations, there are flow models. The first such model was introduced by the French physician Quesnay \cite{quesnay}, who was inspired by the analogy with circulatory flow of the human blood. In this model, there is an economic equilibrium which is stationary, closed and without distinction between productive factors and produced goods. The only source of wealth is agriculture (as appropriate for those times), and its products are then freely distributed among the other social classes.  This approach has been generalized in the 19th century by Fisher, who recognized that the
production cycle of work and goods should be complemented by a second, monetary
circuit. This has been the basis for more recent work (see e.g. \cite{mimkes} and references therein), where Fisher's ideas are cast into thermodynamical language. In particular the production and the money in circle are the analogs of work and heat exchange respectively.

In this work, in the same spirit of modeling economic behavior using general physics, we build a simple hydrodynamic model, based on a leaking bucket analogy introduced in previous works \cite{noi,noi1}, able to reproduce the main conclusions of CA theory in its simplest setup, i.e. a two-country world with country $A$ and country $B$ exchanging two goods within a genuine exchange-based economy and a trade flow ruled only by market forces. Production and exchange costs are introduced in order to evaluate the convenience each country has in producing or sharing a good. Explicit analytic solutions of the dynamical equations are given for different exchange paths (i.e. for different exchange functions). A fixed point analysis is carried out and a region in the parameters' space is found, where the time evolution of the amounts of money for both countries shows a lower bound. This allows us to understand how money affects trade and, in general, to investigate the role of CA law within monetary economies. Finally, advantages and drawbacks of the model are discussed together with perspectives for its extension.

The paper is organized as follows.

In Section 2 we introduce the model system under study within the CA theory framework, i.e. a two-country world with country $A$ and country $B$ exchanging in principle different goods. We adopt a communicating vessels scheme, with each vessel behaving as a leaking bucket \cite{noi}. The exchange of goods between the countries is modeled with a simple superposition of Heavyside theta functions in order to get an analytic solution. In Section 3 some simple examples are discussed in order to illustrate the usefulness of the model. In Section 4 the model is augmented with the introduction of production and exchange cost, in order to be able to evaluate the convenience for each country in producing or sharing a good. Finally, in Section 5 conclusions and perspectives of this work are presented.

\section{The model}

The aim of this Section is to reproduce the main results of classical CA theory in its simplest setup, i.e. a two-country world with country $A$ and country $B$ exchanging in principle different goods.

Suppose two countries $A$ and $B$ are both producers of certain goods, and that they are willing to share their excess production of the latter. More precisely, they agree to share these goods only when their internal demand can be satisfied and a certain reserve is secured. Notice that this is a simplified model, in that it doesn't account for other macroeconomic variables, such as the consumers' different preferences, the role of brand names and product reputations in buyers' decisions, the temporary advantage gained as a consequence of the development of a novel technology, etc. Furthermore in this Section we limit ourselves to a genuine exchange model, without introducing prices in the analysis. This more complicated issue will be addressed in Section 4.

This situation can be modeled, for any of the shared goods, using a communicating vessels scheme \cite{noi}.
\begin{figure}[htb]
\includegraphics[scale=0.7]{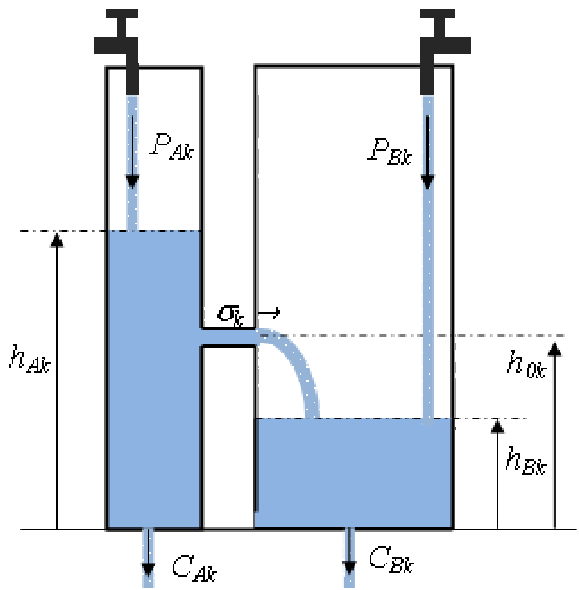}
\caption{Communicating vessel scheme for production, consumption, and exchange of a given good $k$ for countries $A$ and $B$.}
\label{Fig1}
\end{figure}
Since at this level the dynamics related to different goods are decoupled, we consider the case of just one good. In this scheme, the volume of the liquid in each vessel represents the amount of the good that each country holds.
Let us denote by $P_A$, $P_B$ the rates of production of the good in countries $A$ and $B$ respectively, and let $C_A$, $C_B$ be the rates of consumption. Moreover, let $\sigma$ be the rate of exchange. It is assumed that each country starts exchanging after the reserve of the good reaches a given threshold. In the communicating vessel scheme this is denoted by $S_A h_0$ or $S_B h_0$, where $S_{A,B}$ are the sections of the vessels and $h_0$ is a given critical height. Since there is no immediate significance for the sections of the vessels in the economic context, in the following we assume for simplicity that $S_A=S_B=1$.

The dynamics of the system is then given by the following equations
\begin{eqnarray}\label{balanceequations}
P_A-C_A-\sigma F(h_A,h_B)&=&\overset{\cdot}h_A\\\nonumber
P_B-C_B+\sigma F(h_A,h_B)&=&\overset{\cdot}h_B.
\end{eqnarray}
The function $F$ models the exchange between the countries. We choose
\begin{eqnarray}
F(h_A,h_B)&=&\left(\frac{h_A}{h_0}-1\right)\theta\left(\frac{h_A}{h_0}-1\right) \nonumber\\ &&- \left(\frac{h_B}{h_0}-1\right)\theta\left(\frac{h_B}{h_0}-1\right)
\end{eqnarray}
where $\theta(x)$ is the unitary step function, whose r\^{o}le is to ensure that the exchange is turned on only when one of the nations reaches the critical height. Note that this is not an empirical type of law. Rather, with this choice we are assuming that the agreement between the countries is such that the effective exchange rate $\sigma F$ is linear in the excess product if the exchange is unidirectional, and in the difference of excesses if the exchange is bilateral.

We assume consumption rates to be constant. Such a choice is not in line with a strict hydrodynamic model since in that case Torricelli's law would apply, i.e. $C_{A,B}\sim \sqrt{h_{A,B}}$.

Before going on it is convenient to normalize all the variables by defining
\begin{eqnarray}
\tilde{P}_{A,B}&=&\frac{P_{A,B}}{h_0},\\ \tilde{C}_{A,B}&=&\frac{C_{A,B}}{h_0},\\ \tilde{\sigma}&=&\frac{\sigma}{h_0},\\ \eta_{A,B}&=&\frac{h_{A,B}}{h_0}.
\end{eqnarray}
Since in the following there is no risk of confusion, we shall omit the tildes. In what follows, we shall assume that also $P_{A,B}$ and $\sigma_{A,B}$ are constants. In the new variables, the dynamical equations (\ref{balanceequations}) read:
\begin{eqnarray}\label{balanceequationsnorm}
P_A-C_A-\sigma f(\eta_A,\eta_B)&=& \overset{\cdot}\eta_A\\\nonumber
P_B-C_B+\sigma f(\eta_A,\eta_B)&=& \overset{\cdot}\eta_B.
\end{eqnarray}
where $f(\eta_A,\eta_b)=(\eta_a-1)\theta(\eta_A-1)-(\eta_B-1)\theta(\eta_B-1)$.
We can distinguish the following four cases:
\begin{eqnarray}\label{cases}
f(\eta_A,\eta_B)=\begin{cases}
    0 & \text{if } \eta_A,\eta_B< 1\\
    \eta_A-1  & \text{if } \eta_A>1, \eta_B<1\\
    -(\eta_B-1)  & \text{if } \eta_A<1, \eta_B>1\\
    \eta_A-\eta_B   & \text{if } \eta_A>1, \eta_B>1
    \end{cases}
\end{eqnarray}
In each case the equations admit an analytic solution. Let us study them separately:

\paragraph{ $\eta_A,\eta_B< 1$, $f(\eta_A,\eta_B)=0$. }

This is the simplest case since there is no exchange and the two subsystems are decoupled. The equations read:
\begin{eqnarray}
P_A-C_A&=&\overset{\cdot}\eta_A\\
P_B-C_B&=&\overset{\cdot}\eta_B
\end{eqnarray}
and the solution is:
\begin{eqnarray}
\eta_A(t)&=&(P_A-C_A)t +A_1\\\
\eta_B(t)&=&(P_B-C_B)t+B_1,
\end{eqnarray}
where the integration constants $A_1$ and $B_1$ are to be determined at the time where this condition sets in. This solution is valid until  $\eta_A$ or $\eta_B$ reaches the value $1$, where exchange enters the game.

\paragraph{$\eta_A>1,\eta_B< 1$, $f(\eta_A,\eta_B)=\eta_A-1$. }
In this case country $A$, being above the threshold, has started exchanging goods with country $B$, whose dynamics is therefore driven by that of country $A$. The dynamical equations read:
\begin{eqnarray}
P_A-C_A-\sigma(\eta_A-1)&=&\overset{\cdot}\eta_A\\\label{Caseb}
P_B-C_B+\sigma(\eta_A-1)&=&\overset{\cdot}\eta_B.
\end{eqnarray}
The first equation is independent of $B$. It can be solved to yield:
\begin{eqnarray}
\eta_A(t)= \frac{P_A-C_A+\sigma}{\sigma}+A_2e^{-\sigma t}
\end{eqnarray}
By substituting this solution into the second equation (\ref{Caseb}) we get
\begin{eqnarray}
\overset{\cdot}\eta_B=P_B-C_B+P_A-C_A+\sigma A_2e^{-\sigma t}
\end{eqnarray}
whose solution is:
\begin{eqnarray}
\eta_B(t)=(P_B-C_B+P_A-C_A)t-A_2e^{-\sigma t}+B_2
\end{eqnarray}
As in the previous case, the integration constants must be determined in the instant of time where this regime starts.

\paragraph{$\eta_A<1,\eta_B> 1$, $f(\eta_A,\eta_B)=-(\eta_B-1)$. }

The dynamical equations read:
\begin{eqnarray}
P_A-C_A+\sigma(\eta_B-1)&=&\overset{\cdot}\eta_A\\
P_B-C_B-\sigma(\eta_B-1)&=&\overset{\cdot}\eta_B.
\end{eqnarray}
from which it is clear that the solution can be obtained from the one found in the previous case by exchanging $A$ with $B$. This is to be expected since in this case it is country $B$ that is above the threshold and shares its excess goods with country $A$.

\paragraph{$\eta_A>1,\eta_B> 1$, $f(\eta_A,\eta_B)=\eta_A-\eta_B$. }

In this case both countries are above threshold, so that exchange goes in both ways. The dynamical equations read:
\begin{eqnarray}
P_A-C_A-\sigma(\eta_A-\eta_B)&=&\overset{\cdot}\eta_A\\
P_B-C_B+\sigma(\eta_A-\eta_B)&=&\overset{\cdot}\eta_B.
\end{eqnarray}
We can rewrite this system using vector notation, setting
\begin{eqnarray}
\boldsymbol{\eta}=\left( {\begin{array}{c}
   \eta_A \\ \eta_B      \end{array} } \right),\qquad \mathbf{P} =\left( {\begin{array}{c}
   P_A-C_A \\ P_B - C_B     \end{array} } \right)
\end{eqnarray}
and
\begin{eqnarray}
\mathbf{M}=\sigma\left( {\begin{array}{cc}
  1 & -1 \\ -1 & 1      \end{array} } \right)
\end{eqnarray}
so that
\begin{eqnarray}
\overset{\cdot}{\boldsymbol{\eta}}+ \mathbf{M} \boldsymbol{\eta}=\mathbf{P}.
\end{eqnarray}
The matrix $\mathbf{M}$ is singular, thus one eigenvalue is zero, the other eigenvalue being $2\sigma$. The system can be easily solved, but it is useless to write down the explicit solutions.

\section{A simple example}

In order to illustrate the usefulness of the model, in this section we provide a simple example.

Suppose that the two countries agree to maintain the production rates $P_A$ and $P_B$ at high enough levels so that a steady state of equations (\ref{balanceequationsnorm}) is reached under conditions $d$. Still, the two countries would like not to overproduce the shared goods.

Suppose that the production capability of country $A$ is higher than the production capability of country $B$. The two countries may then agree to reach a steady state defined by
\begin{eqnarray}
\eta_A&=&1+\epsilon,\\
\eta_B&=&1
\end{eqnarray}
Then Eqs. (\ref{balanceequationsnorm}) under steady state conditions tell us that
\begin{eqnarray}
P_A&=&C_A + \epsilon \sigma\\
P_B&=&C_B-\epsilon \sigma
\end{eqnarray}
which means that country $A$ can make an effort to produce more than what expected to reach a stationary state with $\eta_A=\eta_B=1$, while country $B$ can slow down its production rate. Of course, in order to compensate for the effort of country $A$, country $B$ can agree to make an effort on the production of a different good. This possibility will be explored in the next section.

From the above example it is clear that each country benefits from trade, also in the presence of a productivity growth in only one country.

\section{Introducing money}

In order to evaluate the convenience the countries have in producing or sharing a good we need to introduce the production and exchange costs. This is the topic of the present Section, which is devoted to a generalization of the previous model. In this way the role of money in shaping trade patterns will be pointed out clearly.

More precisely, we denote by $x_A$ and $x_B$ the production costs of the given good for both countries. We assume that $x_A<x_B$, in order to take into account the fact that country $A$ has more capability of producing the good. Also, let $y_A$ and $y_B$ be the selling prices of the good on the internal markets of the two countries.

Despite international trade is also determined by various endowments of the countries, it remains implicitly solidary. Indeed, within this cooperating solidary countries environment, each country tends to export goods whose production uses more intensively those factors that are relatively more abundant in the country, typical of the comparative advantage. As anticipated in the Introduction, the comparative advantages are determined by the relative abundance of production factors and production technologies (the relative intensity with which production factors are used in the different sectors). Each country will tend to produce intensive goods in the factors of which it is relatively well-equipped. The country where a factor is relatively abundant exports goods whose output is relatively intensive in that factor and, on the contrary, it imports goods that are relatively intense in the relatively low production factor in the country.
This complex dynamics, according to Heckscher-Ohlin's model \cite{modern1}, in international trade leads to a convergence of the relative prices of traded goods. So, there is a direct relationship between relative prices of goods and factor prices, trade also leads to price factor equilibrium. In some respects, similar conclusion also comes from Samuelson's theory of International Trade and Equalization of Factor Prices \cite{samuelson1,samuelson2}.
On the contrary, "\textit{price differences across countries are determined by trade barriers and by a country's specialization in production.}" \cite{hsieh1}.

After the two countries start exchanging the good, these two relative prices converge to a common value $y$, to which also the price of the good in international trade tends. We consider it to be the case that such a convergence has already happened. We assume moreover that international agreements are such that
\begin{eqnarray}
x_A<y<x_B
\end{eqnarray}
i.e. country $B$ finds itself in the situation that the production cost of the good exceeds the cost of the good on the international market. Then country $B$ can decide to stop producing the good and cover its necessities by importing it from country $A$. In this case we are clearly in case $b$, i.e. $\eta_A>1$, $\eta_B<1$. The situation is described by five variables, $P_A$, $P_B$, $\eta_A$, $\eta_B$ and $\sigma$. Moreover there are the fixed parameters $x_A$, $x_B$, $y$, $C_A$ and $C_B$.  Supposing we are at a fixed point, consistently with our hypothesis that all transients are gone, the equations for the quantities $\eta_A$ and $\eta_B$ are
\begin{eqnarray}
P_A-C_A-\sigma(\eta_A-1)&=&0,\label{fixedpointeq1}\\
P_B-C_B+\sigma(\eta_A-1)&=&0.\label{fixedpointeq2}
\end{eqnarray}
These two equations reduce the number of parameters of the problem from 5 to 3. In the following we consider $\eta_A$, $\eta_B$ and $\sigma$ as independent variables (with the constraint $\eta_A>1$, $\eta_B<1$), and $P_A$ and $P_B$ are determined by the above fixed point equations.

Let us denote by $m_A$ and $m_B$ the amounts of money for countries $A$ and $B$. The equations which describe their variations are:
\begin{eqnarray}
\frac{dm_A}{d t}&=&-x_AP_A+y[C_A + \sigma(\eta_A-1)]\\
\frac{dm_B}{d t}&=&-x_BP_B+y[C_B - \sigma(\eta_A-1)].
\end{eqnarray}
The conditions we seek are $\frac{dm_A}{d t}\geq 0$ and $\frac{dm_B}{d t} \geq 0$.
\begin{eqnarray}
-x_AP_A+y[C_A + \sigma(\eta_A-1)]&\geq&0\\
-x_BP_B+y[C_B - \sigma(\eta_A-1)]&\geq&0.
\end{eqnarray}
Solving the fixed point equations (\ref{fixedpointeq1}) and (\ref{fixedpointeq2}) with respect to the $P$'s and substituting in our conditions we get
\begin{eqnarray}
(y-x_A)[C_A + \sigma(\eta_A-1)]&\geq&0\\
(y-x_B)[C_B - \sigma(\eta_A-1)]&\geq&0.
\end{eqnarray}
Since $C_A + \sigma(\eta_A-1)\geq 0 $, and by hypothesis $-x_A+y\geq 0$ also, the first inequality is always satisfied. The second one requires $C_B - \sigma(\eta_A-1)\leq 0 $, since by hypothesis $-x_B+y\leq 0$. On the other hand the left hand side of this inequality is a non negative quantity being nothing but $P_B$, therefore our condition can be satisfied only if
\begin{eqnarray}
C_B = \sigma(\eta_A-1)
\end{eqnarray}
i.e. if $P_B=0$. Thus the only way for country $B$ not to lose money is to completely stop production of the good, importing from country $A$ all the amount consumed. If country $B$ wants to have a positive income, i.e. $\frac{dm_B}{d t} \geq 0$, the only way is to have a second good for which
\begin{eqnarray}
x_{A2}>y_2>x_{B2}
\end{eqnarray}
so that it is country $A$ that imports the second good from country $B$, so that a symmetric situation is achieved, described by the equations
\begin{eqnarray}
\frac{dm_A}{d t}&=&-x_{A1}P_{A1}+y_1[C_{A1} + \sigma_1(\eta_{A1}-1)]\nonumber\\&&- x_{A2}P_{A2}+y_2[C_{A2} - \sigma_2(\eta_{B2}-1)],\\
\frac{dm_B}{d t}&=&-x_{B1}P_{B1}+y_1[C_{B1} - \sigma_1(\eta_{A1}-1)]\nonumber\\&&- x_{B2}P_{B2}+y_2[C_{B2} + \sigma_2(\eta_{B2}-1)].
\end{eqnarray}
for the amounts of money, while the fixed point equations are
\begin{eqnarray}
P_{A1}&=&C_{A1}+\sigma_1(\eta_{A1}-1),\\
P_{A2}&=&C_{A2}-\sigma_2(\eta_{B2}-1),\\
P_{B1}&=&C_{B1}-\sigma_1(\eta_{A1}-1),\\
P_{B2}&=&C_{B2}+\sigma_2(\eta_{B2}-1),
\end{eqnarray}
where the indices $1$ and $2$ refer to the two different goods. Thus the time evolution of the amounts of money for the two countries at the fixed point are described by
\begin{eqnarray}
\frac{dm_A}{d t}&=&(-x_{A1}+y_1)P_{A1} + (-x_{A2}+y_2)P_{A2},\\
\frac{dm_B}{d t}&=&(-x_{B1}+y_1)P_{B1} + (-x_{B2}+y_2)P_{B2}.
\end{eqnarray}
Putting
\begin{eqnarray}
\alpha_1&=&-x_{A1}+y_1>0,\\
\alpha_2&=&-x_{A2}+y_2<0,\\
\beta_1&=&-x_{B1}+y_1<0,\\
\beta_2&=&-x_{B2}+y_2>0.
\end{eqnarray}
the inequalities $\frac{dm_A}{dt}\geq 0$, $\frac{dm_B}{dt}\geq 0$ read
\begin{eqnarray}
\frac{\alpha_1}{-\alpha_2}&\geq& \frac{P_{A2}}{P_{A1}},\\
\frac{\beta_2}{-\beta_1}&\geq& \frac{P_{B1}}{P_{B2}}
\end{eqnarray}
%
Explicitly, the two inequalities read
\begin{eqnarray}\label{inequality1}
\frac{C_{A2}-\sigma_2(\eta_{B2}-1)}{C_{A1}+\sigma_1(\eta_{A1}-1)}&\leq&-\frac{y_1-x_{A1}}{y_2-x_{A2}},\\\label{inequality2}
\frac{C_{B1}-\sigma_1(\eta_{A1}-1)}{C_{B2}+\sigma_2(\eta_{B2}-1)}&\leq&-\frac{y_2-x_{B2}}{y_1-x_{B1}}.
\end{eqnarray}
Actually, in international trade agreements participating countries are required to match their respective trade balances. In the case of a two side agreement the trade balances are equal and opposite, thus the requirement is that they both be equal to zero. In formulas, we require
\begin{eqnarray}
B_{A1}+B_{A2}=-B_{B1}-B_{B2}=0
\end{eqnarray}
where
\begin{eqnarray}
B_{A1}&=&y_1\sigma_1(\eta_{A1}-1),\\
B_{A2}&=&-y_2\sigma_2(\eta_{B2}-1),\\
B_{B1}&=&-y_1\sigma_1(\eta_{A1}-1),\\
B_{B2}&=&y_2\sigma_2(\eta_{B2}-1).
\end{eqnarray}
This requirement translates in the following relation which links the exchange rates of the two goods:
\begin{eqnarray}
\sigma_2=\frac{\eta_{A1}-1}{\eta_{B2}-1}\frac{y_1}{y_2}\sigma_1.
\end{eqnarray}
Substituting this relation in the inequalities (\ref{inequality1}, \ref{inequality2}), we obtain
\begin{eqnarray}\label{inequalities2}
\frac{C_{A2}-\sigma_1(\eta_{A1}-1)\frac{y_1}{y_2}}{C_{A1}+\sigma_1(\eta_{A1}-1)}&\leq&-\frac{y_1-x_{A1}}{y_2-x_{A2}},\\
\frac{C_{B1}-\sigma_1(\eta_{A1}-1)}{C_{B2}+\sigma_1(\eta_{A1}-1)\frac{y_1}{y_2}}&\leq&-\frac{y_2-x_{B2}}{y_1-x_{B1}}.
\end{eqnarray}
which remarkably depend only on two variables, $\sigma_1$ and $\eta_{A1}$. These inequalities must be satisfied together with the conditions $P_{A2}=C_{A2}-\sigma_1(\eta_{A1}-1)\frac{y_1}{y_2}\geq0$, $P_{B1}=C_{B1}-\sigma_1(\eta_{A1}-1)\geq0$. These two inequalities require that the consumption rates of the imported goods in the two countries be not too small and that the exchange rate $\sigma_1$ be not too large. The figure shows the region in the $\sigma_1-\eta_{A1}$ plane where all inequalities are satisfied.
\begin{figure}[htb]
\includegraphics[scale=1]{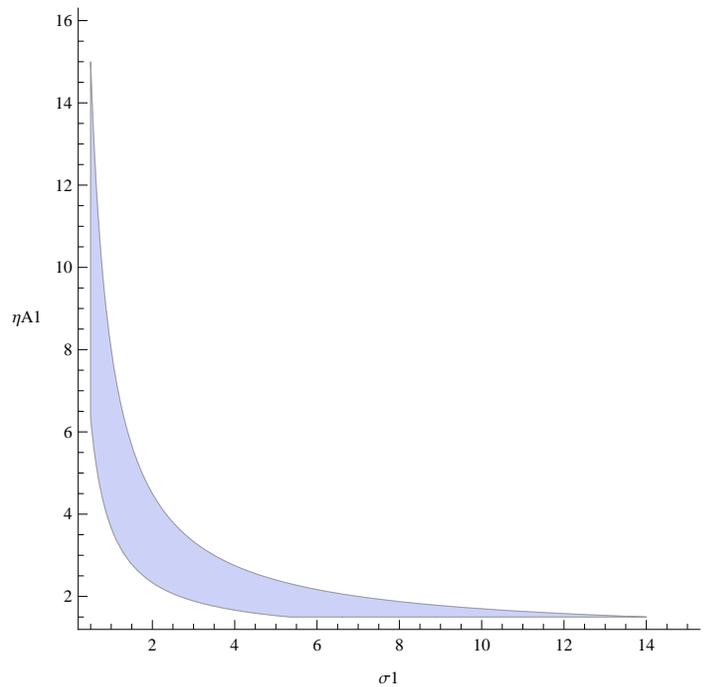}
\caption{Locations of the points where all four inequalities are satisfied. The used values of the parameters are $C_{A1} = 1, C_{A2} = 5, C_{B1} = 7, C_{B2} = 2, x_{A1} = 1, y_1 = 2, x_{B1} = 3, x_{A2} = 5, y_2 = 4, x_{B2} = 2$. We considered $\sigma_1\geq 0.5$ and $\eta_{A1}\geq 1.5$ in order to avoid numerical singularities, but in principle we can have $\sigma_1\geq 0$ and $\eta_{A1}\geq 1$. \label{F1}  }
\end{figure}
As a final remark, our model shows clearly that both countries may benefit from free exchange. Therefore, as a general rule, one may say that, given this occurrence, any trade restriction should be avoided.

\section{Discussion and conclusions}

Although most nations claim to be in favor of free trade, it seems paradoxical that today,  most of them continue to impose many restrictions on international trade. As a result, they advocated restrictions on imports, incentives for exports, and strict government regulation of all economic activities. Indeed, the list of protected products is long and varied. By so doing, both nations end up consuming more of both commodities than without trade.

Despite modern protectionists claim that the comparative cost doctrine has little contemporary validity, and might apply more in a static world in which capital and labor would be fixed in quantity and immobile internationally - and trade restrictions are needed because the economy is not sufficiently adaptable to changing comparative advantages, the international mobility of capital and managerial resources can quickly alter factor proportions, thereby raising the risks of specialization and the costs of adjustment -  the economic lesson of comparative advantage demonstrates, and this work too,  that both international trading partners are best served without trade restrictions. While not denying that comparative advantages  change more rapidly today than in past, the contemporary economy has sufficient flexibility to adjust to such changes.

Comparative advantage shows tariffs and trade quotas protect inefficient firms, harm consumers and lower total productivity. The fact that the country A gains much more than the country B is not important. What is important is that both nations can gain from specialization in production and trade.  With complete specialization, the equilibrium-relative commodity prices will be between the pretrade-relative commodity prices prevailing in each nation.

According to the law of comparative advantage, even if one nation is less efficient than (has an absolute disadvantage with respect to) the other nation in the production of both commodities, there is still a basis for mutually beneficial trade. The less efficient nation should specialize in the production and export of the commodity in which its absolute disadvantage is less. This is the commodity of its comparative advantage. The principle of comparative advantage remains as cogent today as it was in Ricardo's time.

In this work, building on a leaking bucket analogy and a communicating vessels scheme \cite{noi}, we introduced a simple hydrodynamic model, able to reproduce the main conclusions of CA theory in its simplest setup, i.e. a two-country world with country $A$ and country $B$ exchanging two goods within a genuine exchange-based economy and a trade flow ruled only by market forces. We modeled the exchange flux between countries by means of a simple but non trivial exchange function in order to get an analytic solution while retaining all the main phenomenology. Finally, production and exchange costs are included in this framework in order to evaluate the convenience each country has in producing or sharing a good. A fixed point analysis has been carried out and a region in the parameters' space has been found, where the time evolution of the amounts of money for both countries shows a lower bound. That allowed us to assess the role of money in shaping trade patterns.

Our results reproduce the main features of CA theory. In a free exchange world all countries benefit from trade; that happens also in the presence of productivity growth only in one country. More in general, countries gain because they export goods whose prices are relatively higher while import goods whose prices are relatively lower. The focus is on cooperation rather than on competition and on the free exchange flow between countries. We would like to emphasize the r\^{o}le played by the trade balance constraint, which represents the only action required from the government.

We have to point out that our model doesn't take into account the role of other macroeconomic variables, such as the consumers' different preferences, the role of brand names and product reputations in buyers' decisions, the temporary advantage gained as a consequence of the development of a novel technology, etc.. In fact the leaking bucket model is based on the concept of representative agent and, as such, is subjected to all its limitations \cite{DeMartinoMarsili1}. 

Finally, it would be interesting to recast the problem in the language of cooperative Game Theory \cite{GT}. In this context a general $N$-player game could be envisaged with cooperation as the mandatory choice in order to reach a free trade regime.

\section*{Author contribution statement}

All the authors contributed equally to the paper.

\end{document}